# A Multiscale Simulation Approach for Germanium-Hole-Based Quantum Processor

Tong Wu and Jing Guo

**Abstract—** A multiscale simulation method is developed to model a quantum dot (QD) array of germanium (Ge) holes for quantum computing. Guided by three-dimensional numerical quantum device simulations of QD structures, an analytical model of the tunnel coupling between the neighboring hole QDs is obtained. Two-qubit entangling quantum gate operations and quantum circuit characteristics of the QD array processor are then modeled. Device analysis of two-qubit Ge hole quantum gates demonstrates faster gate speed, smaller process variability, and less stringent requirement of feature size, compared to its silicon counterpart. The multiscale simulation method allows assessment of the quantum processor circuit performance from a bottom-up, physics-informed perspective. Application of the simulation method to the Ge QD array processor indicates its promising potential for preparing high-fidelity ansatz states in quantum chemistry simulations.

## I. INTRODUCTION

SEMICONDUCTOR materials provide a promising platform for the hardware realization of quantum computers. Rapid progress on semiconductor-based quantum computers has been achieved recently. In particular, two-qubit quantum gates and quantum processors with fast operation speed and high fidelity have been demonstrated on semiconductors [1][2][3][4][5]. While most experimental demonstrations of semiconductor quantum computing devices are based on electrons, recent experimental demonstrations of hole-based quantum gates and processors show attractive performance potentials [6][7][8]. For example, a single qubit gate fidelity of 99.9899% has been demonstrated for hole spins in germanium (Ge) [9]. In two-qubit quantum gates based on Ge hole spins, a fast two-qubit gate operation time of <10 ns has been achieved [8]. Compared to compound semiconductor materials such as GaAs and InAs, silicon (Si) and Ge can remove nuclear spin dephasing through isotope engineering. These "quieter" semiconductor material systems can help to remove quantum decoherence for achieving longer coherence time and higher quantum fidelity. [10], [11]

Compared to more mature hardware platforms such as superconducting and trapped ion quantum computing [12], semiconductor-based quantum computing is limited to a smaller qubit count. While studies on semiconductor quantum computing have mostly focused on single and two-qubit systems, a $2 \times 2$ quantum dot (QD) array with controllable interdot coupling has previously been demonstrated on GaAs [13]. Recently, a four-qubit quantum processor based on holes in the Ge QD array was successfully demonstrated [8]. Although it only demonstrated semiconductor quantum processors up to 4 qubits, the pioneering work opened a door for scaling up the qubit count of a semiconductor-based quantum processor [8].

Motivated by these recent experiments and the potential of semiconductor QDs for quantum computing in the noisy intermediate-scale quantum (NISQ) era [14], it is imperative to develop computer-aided simulation and design methods for the design of quantum processors based on Ge QD array. While top-down approaches have been generally used for quantum computing algorithms and circuits, and co-design of quantum software and hardware has been reported recently [15], a bottom-up approach that encapsulates essential material and device physics of quantum circuits has not yet been developed. Bottom-up approaches have been shown effective for assessing device options in neuromorphic computing systems [16]. In this study, a multiscale, bottom-up simulation framework is developed to model a quantum processor based on a Ge QD array with a SiGe/Ge heterostructure. We summarize our contributions as follows:

(i) For physical technological computer-aided-design simulation of quantum gate device based on holes in Ge, a three-dimensional numerical device simulation is developed and used to parameterize an analytical model for the tunnel coupling strength between the neighboring QDs.

(ii) We assess the performance potential of a two-qubit quantum gate based on Ge holes and the results show that it can achieve faster gate operation, smaller device-to-device variability, and more relaxed the lithographic size requirement compared to that based on Si holes.

(iii) We develop a bottom-up multiscale simulation method that allows encapsulating physical properties obtained from numerical device simulation into the simulation of quantum circuits for Ge-QD-array-based quantum processors.

(iv) By applying the multiscale simulation method, we show that the Ge QD array processor has the potential to achieve high fidelity in preparing the ansatz state in the variational quantum eigensolver (VQE) algorithm [17].

In the rest of the paper, some related backgrounds and fundamentals of semiconductor quantum computing are discussed in Section II. The structure of the Ge-hole-based quantum computing device and the corresponding modeling

The paper was submitted for review in October 2021. This work is supported by NSF grant #2007200. The authors are with Department of Electrical and Computer Engineering, University of Florida, Gainesville, FL, 32611-6130, USA.

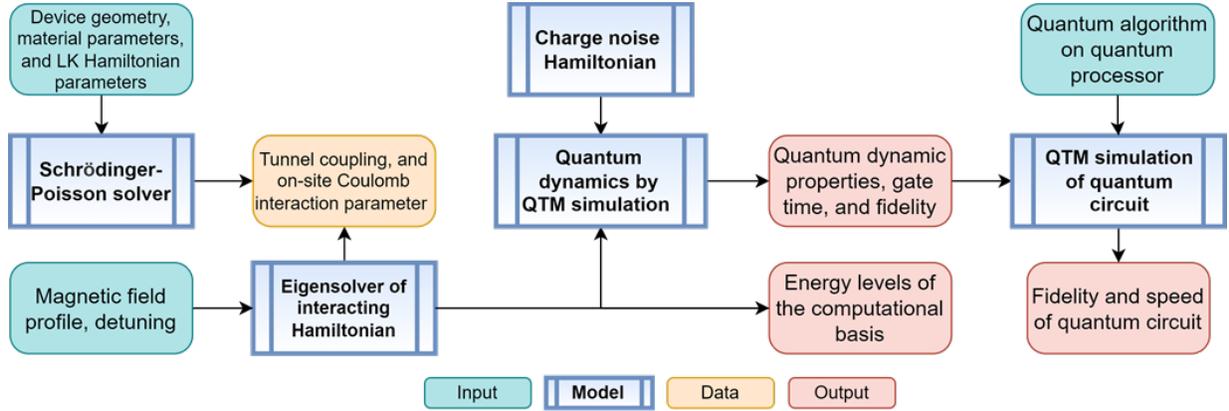

Fig. 1. Overview of the proposed multiscale simulation method for Ge hole-based quantum gates and quantum circuits.

Table. I. Nominal values of the material and device parameters for simulation. (* marks the hitting parameter.)

| Ge Luttinger parameters [25] | | | Device geometry | | | Relative dielectric constant [18], [19], [20] | | | Magnetic field (Zeeman splitting) | | Charge noise* |
|---|---|---|---|---|---|---|---|---|---|---|---|
| $\gamma_1$ | $\gamma_2$ | $\gamma_3$ | Plunger gate | Barrier gate length | Ge thickness | Ge | Si$_{0.2}$Ge$_{0.8}$ | Al$_2$O$_3$ | $E_Z$ | $\Delta E_z$ | $\langle \delta t_c \rangle$ |
| 13.25 | 4.20 | 5.56 | $20 \times 20$ nm$^2$ | $(L_s - 4)$ nm | 20 nm | 16 | 15.2 | 9.8 | 1.0 meV | 0.1 meV | 0.24 μeV |

and simulation are discussed in Section III; the simulation results of QDs and the QD array for quantum processor are presented in Section IV, and the main conclusions are stated in the last section.

## II. Preliminaries

In this section, we briefly review some related backgrounds and fundamentals of semiconductor-based quantum computing. Among various approaches for hardware realization of quantum computing, semiconductor-based approach has the advantage of advanced nanofabrication, excellent scalability, and integratability with integrated circuits. Both nuclear and electron spins, hosted by either semiconductor quantum dots or defect and dopant centers, have been investigated for realizing quantum gates and memory in a variety of semiconductor material systems, including GaAs, Si, Ge, wide-bandgap semiconductors [11].

Compared to compound semiconductor such as GaAs qubits, in group IV semiconductors such as silicon and germanium, especially in isotopically purified group IV semiconductors, where the nuclear spin is nearly zero, spin coherence times can be very long due to weak hyperfine coupling, which is ideal for quantum information processing and storage. Furthermore, silicon and germanium-based quantum computing can harvest and leverage the vast infrastructure and success of the silicon chip industry, which promises compatibility with CMOS technologies, excellent scalability, low fabrication cost, and high integration density. On the other hand, compared to more mature quantum computing hardware platforms such as superconducting quantum computing, the qubit counts of semiconductor quantum chips still lag behind, although the semiconductor approach has excellent potential for scalability [10]. Both electrons and hole spins in semiconductors have been actively explored as quantum information carriers.

Multiscale quantum computer-aided simulation and design can be a powerful tool for exploring the understanding the potential and limitations, and optimizing hardware designs for semiconductor quantum computing.

## III. Multiscale Simulation Approach

A multiscale simulation approach from numerical device simulations to small-scale quantum circuit simulations is developed to describe the operation of a hole qubit array. We make the assumptions that the single qubit gate is ideal, the phase-shifting during the pulsing is calibrated [6], and spin-orbit interaction (SOI) dephasing is neglected. The flowchart of the device simulation is shown in Fig. 1 and described in detail in the subsections below. The nominal values of the simulation parameters used are listed in Table I The multiscale framework is developed for Ge-QD-based quantum gate devices and circuits. It can be extended to quantum gate devices and circuits based on other semiconductor QDs.

### III. A. Modeled Ge Quantum Processor Device Structure

Figure 2(a) shows a schematic top view of the modeled 2×2 QD array, in which the QDs are defined by the plunger gates (PGs) and the barriers between the QDs are modulated by the barrier gates (BGs). In a recent experiment, a QD array based on Ge holes has been demonstrated for four-qubit quantum processor operations [8]. Figure 2(b) shows the cross-sectional view of any pair of neighboring QDs. In the vertical direction, a quantum well is formed in the Ge layer due to heterostructure confinement. The confinement is along the [100] direction of a Ge layer sandwiched by Si$_{0.2}$Ge$_{0.8}$ layers. Vertical confinement of the heterostructure results in the lift of degeneracy between the bulk heavy hole (HH) and light hole (LH) bands as schematically shown in Fig. 2(c). The highest valence subband derived from the bulk HH band hosts hole spins for quantum computing. The schematic subband profile of two neighboring

QDs in the in-plane direction is shown in Fig. 2(d). An entangling two-qubit quantum gate can be achieved by either creating a detuning potential between QDs or by modulating the tunnel coupling [21]. The mechanism of tunnel barrier modulation allows the device to operate at the symmetrically biased points, in which the impact of charge noise can be reduced [22], [23]. We, therefore, focus on modeling tunnel coupling modulation here.

III.B. Device Simulation of Hole-Based Quantum Gate

***Finite-element device simulation:*** The hole QDs for quantum computing are formed on the SiGe/Ge/SiGe heterostructure. Low energetic hole states can be described by the anisotropic Luttinger-Kohn (LK) Hamiltonian approximation [24][25], which is used here for computational efficiency. Atomistic simulations, which are computationally much more demanding, can be useful for describing atomistic scale features of interfaces and defects, which are not treated here. To compute the subband profile and charge density in the heterostructure, we numerically discretize a 4-band LK $k \cdot p$ Hamiltonian, $H_{LK}$, in the vertical confinement direction [25], which treats HH and LH bands as:

$$H_{LK} = \begin{bmatrix} H_{HH} & -S & R & 0 \\ -S^* & H_{LH} & 0 & R \\ R^* & 0 & H_{LH} & S \\ 0 & R^* & S^* & H_{HH} \end{bmatrix}, \quad (1)$$

where

$$H_{HH} = -\frac{\hbar^2 k_z^2}{2m_0}(\gamma_1 - 2\gamma_2) - \frac{\hbar^2(k_x^2+k_y^2)}{2m_0}(\gamma_1 + \gamma_2),$$

$$H_{LH} = -\frac{\hbar^2 k_z^2}{2m_0}(\gamma_1 + 2\gamma_2) - \frac{\hbar^2(k_x^2+k_y^2)}{2m_0}(\gamma_1 - \gamma_2), \quad (2)$$

$$R = -\frac{\sqrt{3}\hbar^2}{2m_0}\left[-\gamma_3(k_x^2 - k_y^2) + 2i\gamma_2 k_x k_y\right],$$

$$S = -\frac{\sqrt{3}\hbar^2}{m_0}\gamma_3(k_x - ik_y)k_z.$$

Here, $\hbar$ is the reduced Planck constant, and the Luttinger parameters that characterize the anisotropic mass of the holes are $\gamma_1 = 13.25, \gamma_2 = 4.20, \gamma_3 = 5.56$ for Ge, and $\gamma_1 = 4.26, \gamma_2 = 0.34, \gamma_3 = 1.45$ for Si [25]. The interface valence band discontinuity is $\Delta E_V \approx 0.3$ eV for a $Si_{1-x}Ge_x/Ge/Si_{1-x}Ge_x$ heterostructure, which corresponds to $x \approx 0.8$. The LK Hamiltonian can be discretized along the z direction to compute the heterostructure charge distribution and subbands of the quantum well. The heavy hole bands behave like a spin-3/2 system, i.e., $j_z = m + s_z = \pm\frac{3}{2}$, where the total z component singular momentum $j_z$ is the sum of the orbital component m and spin component $s_z$.

Strain in the $Si_{1-x}Ge_x/Ge/Si_{1-x}Ge_x$ structure induces a Bir-Pikus (BP) Hamiltonian term $H_{BP}$, which results in a total Hamiltonian of $H = H_{LK} + H_{BP}$. The BP Hamiltonian can be approximated as adding additional diagonal terms to the corresponding HH and LH terms in (2), with [26]

$$H_{HH,\epsilon} = -a_v(\epsilon_{xx} + \epsilon_{yy} + \epsilon_{zz}), \quad (3)$$
$$H_{LH,\epsilon} = -\frac{b_v}{2}(\epsilon_{xx} + \epsilon_{yy} - 2\epsilon_{zz}), \quad (4)$$

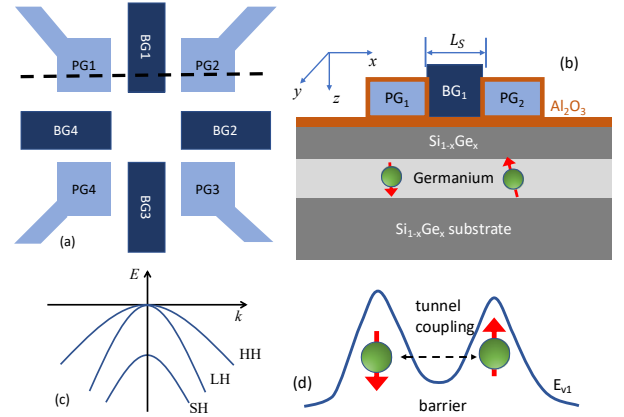

Fig. 2. (a) Schematic layout of a 2 × 2 QD array for a quantum processor. $PG_i$ is the *i*th plunger gate, and $BG_i$ is the *i*th barrier gate. (b) The schematic cross section between 2 neighboring QDs cut at the dashed line in (a). (c) Schematic bulk hole band structure with HH, LH, and split-off hole (SH) bands. (d) Along $x$ direction, the plunger gates are two hole QDs, and the barrier gate modulates the tunnel barrier between QDs.

where $a_v \approx 2.0$ eV and $b_v = -2.3$ eV. By taking the strain values of $\epsilon_{xx} = \epsilon_{yy} \approx -0.006$ and $\epsilon_{zz} \approx 0.0042$ [26], the major effect of the BP Hamiltonian is to change the HH-LH energy split by a constant of ~40 meV.

To simulate the quantum gate device shown in Fig. 2(b), numerical device simulations are first performed by solving a 3-D Poisson equation with the Schrödinger equation by using the finite element method (FEM). The FEM 3-D Poisson solver is discretized by using a prism lattice for the device [27]. The Schrödinger equation is solved by discretizing the LK Hamiltonian for holes in the finite element grid. A mode space approach, which first computes the vertical confinement modes and uses the mode space as the basis in the vertical direction with the discretization of the Hamiltonian in real space for the horizontal space, is used to expedite the solution of the Schrödinger equation [28][29]. The single-particle wave equation and eigen-energies obtained from the FEM Schrödinger-Poisson device simulation can be subsequently used to parameterize the tunnel coupling and on-site Coulomb repulsion terms in the quantum gate Hamiltonian as described below.

***Quantum gate Hamiltonian and parameter extraction from device simulation:*** The Hamiltonian of an entangling quantum gate between two neighboring QDs, as shown in Figs. 2(b) and (d), can be described as follows in the basis set of the HH states $\{|\uparrow\uparrow\rangle, |\uparrow\downarrow\rangle, |\downarrow\uparrow\rangle, |\downarrow\downarrow\rangle, S_{20}, S_{02}\}$ [30],

$$H_0 = \begin{bmatrix} E_z/2 & 0 & 0 & 0 & 0 & 0 \\ 0 & \Delta E_z/2 & 0 & 0 & t_c & t_c \\ 0 & 0 & -\Delta E_z/2 & 0 & -t_c & -t_c \\ 0 & 0 & 0 & -E_z/2 & 0 & 0 \\ 0 & t_c & -t_c & 0 & U_1 - \epsilon & 0 \\ 0 & t_c & -t_c & 0 & 0 & U_2 + \epsilon \end{bmatrix}, \quad (5)$$

where $S_{20}$ ($S_{02}$) are doubly occupied singlet states on QD1 (QD2), the Zeeman splits are $E_z \approx \mu_B(g_1 B_1 + g_2 B_2)$, $\Delta E_z \approx \mu_B(g_1 B_1 - g_2 B_2)$, in which $\mu_B$ is the Bohr magneton, $B_{1,2}$ are the static magnetic fields and $g_{1,2}$ are the effective Ge HH g-factors [31] of the QDs 1 and 2 respectively, $t_c$ is the tunnel coupling, $U_{1,2}$ is the on-site Coulomb interaction, $\epsilon =$ $-q(V_{PG1} - V_{PG2})$, where q is the elementary electron charge, is the detuning energy controlled by the applied detuning voltage, which is set to $\epsilon = 0$ for the symmetrically biased point.

To understand the quantum gate operation on the computational basis, the Hamiltonian can be projected to the computational basis of $\{|\uparrow\uparrow\rangle, |\uparrow\downarrow\rangle, |\downarrow\uparrow\rangle, |\downarrow\downarrow\rangle\}$ by using the Schrieffer-Wolff transformation. The effective Hamiltonian can be expressed as [30],

$$H_{eff} \approx \mu_B(g_1 B_1 s_{1,Z} + g_2 B_2 s_{2,Z}) + J\left(\mathbf{s_1} \cdot \mathbf{s_2} - \frac{1}{4}\right)$$

$$\approx \begin{bmatrix} E_z/2 & 0 & 0 & 0 \\ 0 & \Delta E_z/2 - J/2 & J/2 & 0 \\ 0 & J/2 & -\Delta E_z/2 - J/2 & 0 \\ 0 & 0 & 0 & -E_z/2 \end{bmatrix} \quad (6)$$

where $s_{1,2}$ are the spin-1/2 operator on QD1 (QD2), the subscript z denotes its z component, and the exchange interaction can be expressed as,

$$J \approx \frac{2t_c^2(U_1+U_2)}{(U_1-\epsilon)(U_2+\epsilon)}. \quad (7)$$

In the adiabatic approximation, the off-diagonal terms of $H_{eff}$ renormalize the states of $\{|\uparrow\downarrow\rangle, |\downarrow\uparrow\rangle\}$ to a new basis set $\{|\widetilde{\uparrow\downarrow}\rangle, |\widetilde{\downarrow\uparrow}\rangle\}$. By modulating the exchange J through controlling the tunnel barrier height, and thereby, the tunnel coupling, the energy shift $-J/2$ of $\{|\widetilde{\uparrow\downarrow}\rangle, |\widetilde{\downarrow\uparrow}\rangle\}$ with regard to $\{|\uparrow\uparrow\rangle, |\downarrow\downarrow\rangle\}$ can be used to create two-qubit quantum gates. [8] The operation difference between the controlled-phase and controlled-Z (CZ) quantum gates is only single qubit operations [8], [21]. At the symmetrically biased case, i.e., $\epsilon = 0$, modulation of exchange is achieved through modulation of the tunnel coupling by the barrier gate voltage. It is important to model the dependence of the tunnel coupling on the barrier gate voltage accurately.

The value of the tunnel coupling between the QDs can be simulated numerically from the energy levels obtained by using the FEM device simulations described earlier. For the symmetric double quantum dot (DQD) structure, the lowest eigen-energy state is a binding state with energy $E_B$ and the 2nd lowest energy state is an anti-binding state with energy $E_{AB}$. [32] The tunnel coupling matrix element can be expressed as $t_c \approx |\langle \psi_1(\vec{r})|\hat{H}(\vec{r})|\psi_2(\vec{r})\rangle|$, where $\psi_{1(2)}$ is the ground-state wave function of QD1 (QD2) and $\hat{H}$ is the Hamiltonian of the DQD system. In the weak tunneling regime, $\langle \psi_1|\psi_2\rangle \ll 1$, the binding state wave function is $\phi_B \approx \frac{1}{\sqrt{2}}(\psi_1 + \psi_2)$, and the anti-binding state wave function is $\phi_{AB} \approx \frac{1}{\sqrt{2}}(\psi_1 - \psi_2)$. The energy difference between the anti-binding state and the binding state is $|E_{AB} - E_B| \approx |\langle \phi_{AB}|\hat{H}|\phi_{AB}\rangle - \langle \phi_B|\hat{H}|\phi_B\rangle| \approx 2t_c$. The tunnel coupling can be computed by numerically simulated $E_{AB}$ and $E_B$,

$$t_c = |E_{AB} - E_B|/2, \quad (8)$$

. The on-site Coulomb interaction is calculated by numerical integration as:

$$U = \frac{q^2}{4\pi\epsilon_0\epsilon_r} \int d\vec{r}_1 d\vec{r}_2 \frac{|\psi(\vec{r}_1)|^2 |\psi(\vec{r}_2)|^2}{|\vec{r}_1-\vec{r}_2|}, \quad (9)$$

where $\epsilon_0$ and $\epsilon_r$ are the vacuum and relative dielectric constant, respectively, and $\psi(\vec{r})$ is the QD wave function when one dot is occupied.

III.C. Modeling of Dephasing Noise and Quantum Gate Dynamics

*Charge noise modeling:* At the symmetrically biased point, the tunnel noise is dominant over the detuning noise, which is modeled in [33]. The tunnel noise Hamiltonian in the basis of $\{|\uparrow\downarrow\rangle, |\downarrow\uparrow\rangle, S_{20}, S_{02}\}$ can be expressed as [34],

$$H_n = \begin{bmatrix} 0 & 0 & 1 & 1 \\ 0 & 0 & -1 & -1 \\ 1 & -1 & 0 & 0 \\ 1 & -1 & 0 & 0 \end{bmatrix} \delta t_c, \quad (10)$$

where $\delta t_c$ is the stochastic fluctuation of the tunnel coupling due to charge noise. $\delta t_c$ is assumed to follow a Gaussian distribution with a mean value of 0 and the standard deviation of $A_n$, which characterizes the noise amplitude. In the time domain, the noise is assumed to obey the stochastic time dynamics of random telegraph noise [35], with a characteristic time of $\tau_n$. Due to the small size of the QD, only a single or small number of two-level fluctuators are expected for each qubit [22]. Furthermore, the characteristic time of the charge traps is much longer than the quantum gate time of nanoseconds, and the charge noise spectrum density is dominantly in the low-frequency range [36]. The results are insensitive to the exact noise spectral distribution.

*Noise due to nuclear spin and spin-orbit interaction:* In this study, hole spin dephasing due to nuclear spin is assumed to be neglected, because the nuclear spin noise in Ge can be removed by using isotopically purified Ge.

For Ge hole spin qubits, all-electric-control of single-qubit gates can be achieved based on electric dipole spin resonance (EDSR) [37]. Despite noise due to spin-orbit interaction, single qubit gates with very high fidelity values have been demonstrated [9]. In this work, we focus on two-qubit quantum gates, where charge noise is dominant and whose fidelity is limiting the overall quantum circuit performance, and neglect noise of single-qubit gates.

*Quantum gate dynamics and performance assessment:* A quantum trajectories method (QTM) is used to simulate the time-evolution of quantum gates and quantum circuits in the presence of noise. The quantum gate operator is calculated by exponentiating the time integral of the Hamiltonian, averaged over the stochastic realizations of quantum trajectories in Monte-Carlo sampling of the charge noise Hamiltonian [23] [38]. Compared to directly solving the master equation, QTM reduces the computation of evolution of $O(N^2)$ density matrix elements to simpler Monte Carlo simulation of $O(N)$ quantum state space, which helps to improve the time efficiency of simulation. Furthermore, the QTM method treats non-Markovian evolution of the open-quantum system [38]. The method is applied to simulate quantum dynamic properties of both the two-qubit quantum gates and quantum processor circuits consisting of multiple qubits as described in detail below in Section III.D. The quantum gate time is measured for the CZ quantum gate, with the phase of the target qubit rotated by $\pi$.

III.D. Simulation of QD Array Processor

To model the implementation of a quantum algorithm on a quantum circuit based on Ge QD array, we straightforwardly evolve the many-body wave function in the $2^N$-dimensional Fock space by cascading one-qubit and two-qubit quantum gates. A single qubit fidelity of 99.9899% has been demonstrated for Ge hole spins[9]. Because of significantly higher fidelity and faster gate operation compared to the two-qubit entangling gate operations, the single-qubit quantum gates are not the limiting factor of the quantum circuit fidelity. The single-qubit operations are assumed to be ideal. For each two-qubit gate operation, the charge noise model described in (10) is used. The result, therefore, represents the upper limit of the quantum circuit fidelity limited by charge-noise of the two-qubit quantum gate operations.

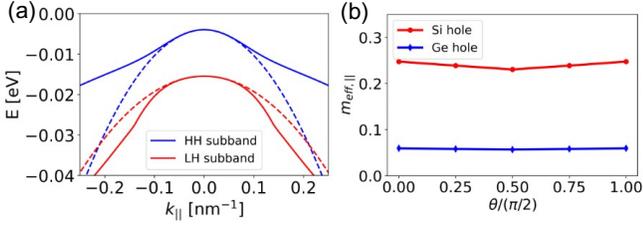

Fig. 3. (a) The simulated in-plane E-k relation (solid lines) of the HH and LH subbands for the $Si_{1-x}Ge_x/Ge/Si_{1-x}Ge_x$ heterostructure. The dashed lines show parabolic fitting to extract the in-plane effective mass values. (b) The extracted in-plane effective mass vs. crystal direction of Ge, compared to Si, where $\theta$ is the angle to the horizontal [100] crystal direction. The Ge or Si layer thickness is 20 nm.

To model dephasing and noise in preparing a quantum state by using a quantum processor, the QTM is used to stochastically evolve the many-body initial quantum state of multiple qubits through Monte Carlo sampling of quantum trajectories [38]. The stochastic Schrödinger equations in the presence of stochastic noise are solved to determine time-dependent evolution of quantum states (quantum trajectories),

$$\psi_{n,k}(t) = exp\left(-\frac{i}{\hbar}(H_0 + H_{n,k})t\right)\psi_0, \quad (11)$$

where $\psi_0$ is the initial wave state and $H_{n,k}$ is the kth Monte-Carlo realization of the noise Hamiltonian. The method allows the system to be modeled by Monte Carlo simulations of quantum wave states instead of handling a matrix equation [38]. The physical quantities of interest can be obtained from expectation values of Monte Carlo samples of the quantum trajectories. The fidelity of preparing the quantum state is assessed as:

$$\mathcal{F} = \langle\psi_{f,ideal}|\rho_{f,n}|\psi_{f,ideal}\rangle, \quad (12)$$

where $\psi_{f,ideal}$ is the ideal wave function, and the noisy density matrix $\rho_{f,n} = \overline{|\psi_{f,n}\rangle\langle\psi_{f,n}|}\,\psi_{f,n}$ is the noisy wave function of the final state, and the average is over quantum trajectories.

III.E. Limitations of the Multiscale Simulation Method for Ge-QD Quantum Processor

The multiscale simulation method provides a framework to evaluate Ge-based quantum processors with essential material and device physics encapsulated into quantum circuit simulations. The capability is especially important for physical designs of the quantum processor. It, however, still has the following limitations:
(i) Although the multiscale approach already significantly reduced computational cost for bottom-up quantum circuit simulations, the simulation is performed in a classical computer, which limits the simulation to a quantum processor with a relatively small number of qubit counts.

(ii) The coupling between the quantum dots in the circuit is limited to nearest neighbors. The nearest neighbor coupling is what was used in state-of-the-art experiments to demonstrate simple algorithms [8].

Limitation (i) is imposed by the fact that the computational cost to simulate an entangled state on a classical computer exponentially grows as the qubit counts increase. For limitation (ii), an interconnection scheme such as exchange-based quantum state transfer only requires nearest neighbor coupling [39]. A resonator-based interconnecting scheme for remote coupling of semiconductor-based qubits has also been experimentally explored [40]. Future studies are needed to systematically study and model these quantum interconnection schemes. State-of-the-art experimental demonstrations of semiconductor-QD-based processors have been limited to four qubits with nearest neighbor coupling. Considering the limitations, the multiscale simulation method is intended for early physical design and exploration of semiconductor-QD based processors to discuss device impacts on small-scale quantum processor characteristics and performance limits by quantum hardware engineering.

IV. RESULTS FROM MULTISCALE SIMULATIONS OF GE-QD-BASED QUANTUM PROCESSORS

IV.A. Heterostructure Simulation for In-Plane Effective Mass

The tunnel coupling between neighboring quantum dots is strongly dependent on the tunneling effective mass. We first examine the horizontal in-plane E-k relation. Figure 3(a) and (b) show the E-k relation of the highest-HH and LH subbands along the in-plane [100] direction. The large subband spacing of the Ge layer is beneficial for suppressing of decoherence of HH spins due to SOI [37]. The extracted in-plane effective mass of HHs is lighter than that of the LHs, which is referred as mass reversal. Figure 3(b) plots the in-plane effective mass values for Ge and Si structures. The results show that the in-plane effective mass is approximately independent of the in-plane direction. Furthermore, the very light effective mass of Ge, $m^*_{||,Ge} \approx 0.058$, is nearly 4 times smaller than that of Si, $m^*_{||,Si} \approx 0.24$. In addition, we also vary the semiconductor layer thickness between 5 nm and 20 nm, and the extracted in-plane effective mass values remains approximately the same. The much smaller effective mass can significantly enhance the tunneling coupling and quantum entanglement between neighboring spins of the DQD structure for two-qubit quantum gates, as discussed below.

IV. B. Two-qubit Quantum Gate Simulation and Model Parameter Extraction

In a 2D array structure for the QD-based quantum processor as shown in Fig. 2(a), two-qubit quantum gate operations can be realized between any pairs of neighboring QDs, whose cross section is shown in Fig. 2(b). Figure 4(a) shows the simulated HH subband profile along the DQD direction (x-direction in Fig. 2(b)). The horizontal dashed line shows the simulated ground state of the DQD structure. The corresponding wave function is shown in Fig. 4(b), which is a symmetric binding state between two QDs. The wave function of the next state is shown in Fig. 4(c), which is an anti-binding state. The behavior resembles an H$_2$ molecule, and the tunnel coupling determines the energy spacing between the biding and anti-binding states.

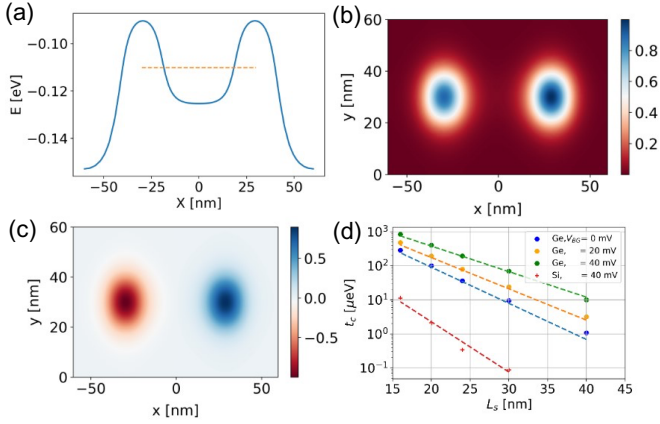

Fig. 4. (a) The valence subband profile along *x* for the two-qubit hole quantum gate as shown in Fig. 2(b), where $L_S = 40$ nm and $|V_{BG}| = 40$ mV. The Si$_{0.2}$Ge$_{0.8}$ top layer thickness is 10 nm. The simulated wave function of (b) binding state and (c) anti-binding state. (d) Tunnel coupling vs. the DQD spacing at different barrier gate voltage magnitudes of $|V_{BG}| = 0, 20, 40$ mV. The lines show the analytical model, and the dots show the numerical simulation results.

Electrostatic gate modulation of the tunnel coupling plays a central role in two-qubit quantum gates. We next examine its dependence on the barrier gate voltage and DQD spacing. The symbols in Fig. 4(d) show the simulated tunnel coupling vs. the dot spacing L$_S$ shown in Fig. 2(b), at different barrier gate voltages. We focus on the gate voltage choice and DQD designs that produce a tunnel coupling value t$_c$ around the range of ~1 μeV to ~100 μeV. Furthermore, the applied barrier gate voltage range shall not produce an excessively low barrier so that the two QDs are not well confined. The exponential sensitivity of the tunnel coupling to the DQD spacing and applied voltage is a signature of quantum tunneling behavior.

To enable efficient simulations of quantum processor, the above numerical simulation of tunneling coupling between neighboring QDs in the processor is parameterized to a physics-based analytical model. The tunnel coupling can be described by an analytical model from the WKB approximation, which is shown by the lines in Fig. 4(d). In the model, the tunnel coupling can be approximated as [41],

$$t_c = t_0 \exp\left(-\frac{\sqrt{2m^* E_b}}{\hbar} L_S\right), \quad (13)$$

where t$_0$ is a tunnel coupling parameter independent of E$_b$ and L$_S$ but dependent on the material such as Si or Ge, m$^*$ is the in-plane effective mass, and the barrier height is

$$E_b = E_{b0} - \beta q V_{BG}, \quad (14)$$

where β is the gating efficiency factor of the barrier gate, E$_{b0}$ is the barrier height constant, and V$_{BG}$ is the barrier gate voltage. The model is fitted to the numerical simulation results in Fig. 4(d) with the fitting values of $\beta = 0.5$, $E_{b0} = 40$ meV, and $t_0 = 12$ meV and $t_0 = 2$ meV for Ge and Si, respectively.

Figure 5 plots the exchange interaction vs. the magnitude of the barrier gate voltage, compared between Ge hole two-qubit gates with different DQD spacing and a silicon two-qubit gate. The barrier gate effectively modulates the tunnel coupling, leading to an average inverse slope of ~20 mV/dec modulation of the tunnel coupling for Ge at a DQD spacing of $L_s = 30$ nm. The slope is even steeper as L$_s$ increases, which also leads to a decrease of the exchange interaction. Benefiting from the smaller hole effective mass, Ge achieves a similar switching behavior compared to a silicon hole-based quantum gate with significantly shorter DQD spacing, which is illustrated by the case of $L_s = 40$ nm for Ge compared to $L_s = 16$ nm for Si as shown in Fig. 5, with a slightly steeper switching slope.

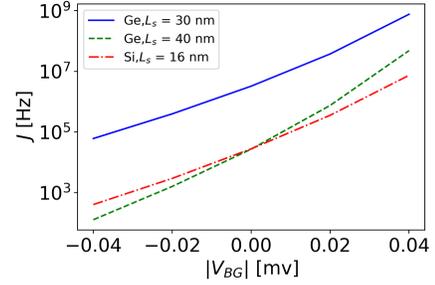

Fig. 5. Exchange interaction *J* vs. the barrier gate voltage magnitude for the Ge hole gate with a DQD spacing of $L_S = 30$ nm (blue solid curve) and 40 nm (green dash curve) and a Si hole gate with $L_S = 16$ nm (red dash-dot curve).

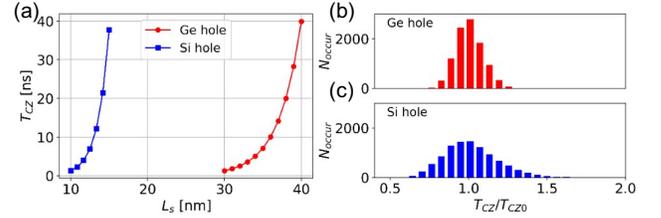

Fig. 6. Quantum gate speed and variability: (a) the gate time vs. the DQD spacing $L_S$ for Ge- and Si-hole-based CZ gates. The magnitude of the applied barrier gate voltage is fixed at $|V_{BG}| = 40$ mV. (b) Histogram distribution of the normalized CZ gate time for Ge hole, compared to (c) Si hole, with the DQD spacing $L_S$ variations, which has a Gaussian distribution with a standard deviation of 0.5 nm. The distribution of the normalized gate time is insensitive to the exact value of $L_S$. The normalized standard deviation values are $\delta T_{CZ}/T_{CZ0} = 0.087$ and 0.173 for Ge and Si holes, respectively, where $T_{CZ0}$ is the nominal value of the gate time without variation.

We next quantify the improvement of the two-qubit quantum gate speed of Ge holes, benefiting from the small hole effective mass and enhanced tunnel coupling. Figure 6(a) shows the quantum gate time T$_{CZ}$ as a function of the device size. To achieve a fast, sub-10 ns CZ gate time, a DQD spacing of $L_s < 37$ nm is needed for Ge holes. However, the requirement is $L_s < 13$ nm for Si holes, which is nearly 3 times more stringent. The comparison is done at a similar barrier height between DQDs. The gate time is sensitive to the exchange coupling determined by the tunnel coupling strength. The smaller effective mass of Ge holes facilitates tunnel coupling for faster gate operations. The lithographic feature size requirement is much less stringent in the Ge system.

To scale up the qubit count in an integrated quantum system, device-to-device variabilities impose significant challenges. Semiconductor fabrication process variability can result in variations in the QD spacing. The effect can be especially important for QD-based quantum processors because the tunnel coupling, which determines the strength of entanglement between neighboring dots, is exponentially sensitive to the QD spacing. To explore the device variability, Monte-Carlo simulations are performed to sample a Gaussian distribution of the DQD spacing, and the simulated distributions of the normalized CZ gate time of Ge and Si are shown in Figs. 6(b) and (c). The results can be understood by differentiating (13),

$$\frac{1}{t_c}\left|\frac{dt_c}{dL_s}\right| = \frac{\sqrt{2m^*E_b}}{\hbar}, \quad (15)$$

which indicates that the normalized sensitivity of $t_c$ to $L_s$ is proportional to $\sqrt{m^*}$. The smaller effective mass of Ge holes results in reduced process-induced device-to-device variability.

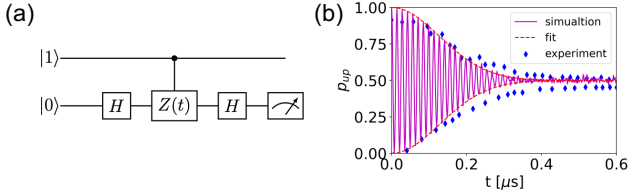

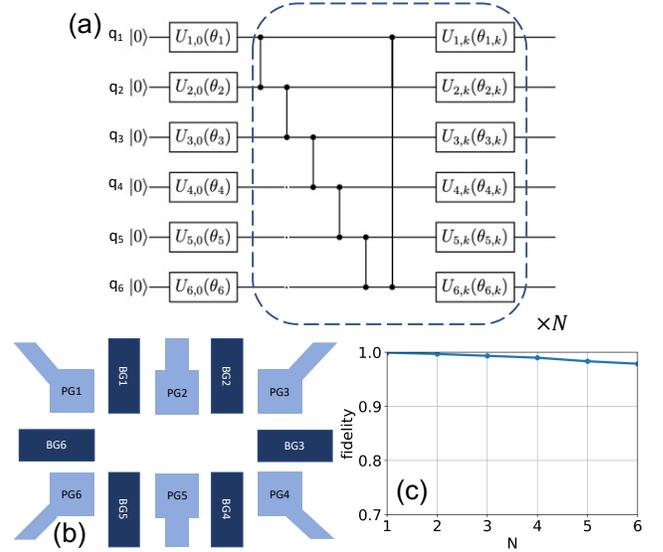

Fig. 7. (a) Quantum circuit diagram for simulating exchange oscillation in (b). The controlled phase gate $Z(t)$ is realized by the DQD device structure and simulated using the QTM as described in text. (b) The simulated exchange oscillation (solid line) compared to the experimental data of the exchange oscillation envelope extracted from Ref. [8] (symbols). The dashed line is an envelope function of $p_{up} = \exp(-(t/\tau)^2)$ fitted to the simulated oscillation with $\tau = 180\ ns$. $p_{up}$ is the spin up probability.

Quantum dynamic characteristics of the two-qubit Ge hole quantum gate in the presence of charge noise are investigated next. We simulate the exchange oscillations in the two-qubit CPHASE gate by using the QTM, as shown in Fig. 7(a). The modeled Ge device structure has a DQD spacing of $L_s = 35$ nm and a plunger gate size of $20 \times 20$ nm$^2$, which results in a tunnel coupling of $t_c \approx 28.4$ μeV and on-site Coulomb interaction of $U_1 \approx U_2 \approx 11$ meV. The QTM simulation results, as shown in Fig. 7(b), capture non-Markovian dynamics of exchange oscillation, in which the envelope of oscillation fits to $exp(-(t/\tau)^2)$. The expression is a Kohlrausch-Williams-Watts (KWW) function with an exponential component $\gamma = 2$, which takes the non-Markovian feature of the Rabi oscillation decay into consideration and is previously used to extract spin dephasing time from experimental data [42]. Furthermore, by comparing to the experimental data from [8], the phenomenological charge noise magnitude can be extracted as $\langle \delta t_c \rangle \approx 0.24$ μeV, which results in a decay time of $\tau \approx 180$ ns. The long decay time compared to the CZ gate time promises high gate fidelity.

IV.C. Model Ge QD array for Quantum Processor

A recent experiment has demonstrated the generation of a four-qubit entangling Greenberger−Horne−Zeilinger (GHZ) state on a 2×2 QD array processor [8]. In a QD array processor, a universal set of quantum gates for quantum computing can be realized by two-qubit gates discussed above together with one-qubit rotational gates. Quantum chemistry simulation of a small molecule can provide a concrete context to explore and assess the potential practical applications of the QD array processor as a noisy intermediate-scale quantum (NISQ) hardware [17][43].

Fig. 8. (a) Quantum circuit for preparing a VQE ansatz state in simulation a BeH$_2$ molecule. The subcircuit in the cashed box can be repeated in cascade for $N \geq 1$ times. (b) Design of a six-qubit quantum processor of a 2D QD array for efficiently implementing the quantum circuit in (a). (c) The fidelity of preparing the ansatz

We next explore the design of a QD array processor for the VQE algorithm, which is a widely used algorithm in quantum chemistry [17][43]. A key challenge for quantum chemistry simulation is to achieve chemical accuracy, for which high fidelity of preparing the ansatz state is indispensable. As an example of VQE simulation, six qubits are sufficient for preparing the ansatz state for simulating the BeH$_2$ molecule, with a quantum circuit as shown in Fig. 8(a) [17][44]. The subcircuit in the dashed line box needs to be repeated in cascade for N times to provide flexibility of parameterizing the ansatz state. This ansatz circuit can be efficiently implemented with a 2×3 QD array, whose schematic top view is shown in Fig. 8(b). The six qubits in the quantum circuit in Fig. 8(a) reside in the QDs defined by PG1 to PG6 in order. For each CZ gate, it only involves two nearest neighbors in the QD array as shown in Table. I, which can be achieved by modulating the barrier gate between the pair of QDs. The designed QD array, therefore, can provide an efficient platform for preparing the ansatz state in VQE simulation of a small molecule.

To assess the fidelity for preparing the ansatz state, we simulate the quantum circuit by considering the decoherence stemming from each two-qubit quantum gate operation, by using the CZ gate device and noise parameters extracted in Fig. 8. The assessment, therefore, represents the upper limit due to the charge noise in each two-quantum gate operation. For the VQE circuit parameters used to prepare the ansatz state of BeH$_2$, we simulate the fidelity of preparing the quantum ansatz state as a function of the number of cascade stages N, as shown in Fig. 8(c). The results show the potential to achieve an ansatz state preparation fidelity of $F > 0.99$ when the circuit is shallow with $N = 1$. To achieve better flexibility of the ansatz state, a larger value of N is often required, which results in a deeper quantum circuit with decreased fidelity. Even with the circuit cascade depth increasing to $N = 6$, the prepared ansatz state fidelity is $F > 0.96$. The high fidelity is due to the fast, sub-10 ns CZ gate operation compared to the quantum decoherence time. Furthermore, the non-Markovian dynamics results in a slower initial decay of $\sim exp\left(-(t/\tau)^2\right)$ for fidelity, compared to the simple Markovian exponential decay of $\sim exp\left(-t/\tau'\right)$. The results indicate the potential of Ge-hole-based QD array processors in implementing the VQE algorithm for quantum chemistry simulations.

## V. Conclusions

A multiscale simulation method is developed to model and assess the Ge-hole-based QD array for quantum processor. The multiscale process takes a bottom-up approach, which allows essential device physics to be incorporated in the assessment of quantum circuit performance for a Ge-QD-based quantum processor. The results show that the Ge hole array provides a promising semiconductor platform to enhance entanglement between neighboring QDs for two-qubit quantum gate noise. Furthermore, a two-qubit quantum gate based on holes in Ge can achieve fast gate speed, and smaller device variability compared to its Si counterpart. To efficiently simulate the QD array for implementing a quantum circuit in the quantum processor, a simple analytical model is extracted from numerical quantum device simulations to describe the dependence of the tunnel coupling on the applied gate voltage and device size. Design and multiscale simulation of the Ge QD array processor shows its potential to achieve high fidelity in preparing the ansatz state of quantum chemistry simulations based on VQE. The bottom-up, multiscale method developed here can allow physical design and assessment of semiconductor-QD-based quantum processors from physical properties of quantum gate devices and their underlying material properties.